\documentclass[sigconf,nonacm]{acmart}

\usepackage[official]{eurosym}
\usepackage[greek,ngerman,english]{babel} 
\usepackage[utf8]{inputenc}
\usepackage{textgreek}
\usepackage{parskip}
\usepackage{pdfpages}
\usepackage{url}
\usepackage{enumitem}

\usepackage{hyperref}

\usepackage{subcaption}
\usepackage{acronym}

\usepackage{graphicx}
\usepackage{amsmath}
\usepackage{float}
\floatstyle{plaintop}
\restylefloat{table}
\restylefloat{figure}
\usepackage{caption} 
\usepackage{blindtext}

\usepackage{booktabs}
\usepackage{eurosym}
\usepackage{amstext} %
\DeclareRobustCommand{\officialeuro}{%
  \ifmmode\expandafter\text\fi
  {\fontencoding{U}\fontfamily{eurosym}\selectfont e}}

\usepackage{acronym}

\AtBeginDocument{%
  \providecommand\BibTeX{{%
    \normalfont B\kern-0.5em{\scshape i\kern-0.25em b}\kern-0.8em\TeX}}}

\begin{document}

\title{Predicting Companies' ESG Ratings from News Articles Using Multivariate Timeseries Analysis}

\author{Tanja Aue}
\affiliation{%
  \institution{University of Innsbruck}
  \city{Innsbruck}
  \country{Austria}}
\email{tanja.aue@student.uibk.ac.at}

\author{Adam Jatowt}
\affiliation{
  \institution{University of Innsbruck}
  \city{Innsbruck}
  \country{Austria}}
\email{adam.jatowt@uibk.ac.at}

\author{Michael Färber}
\affiliation{
  \institution{Karlsruhe Institute of Technology}
  \city{Karlsruhe}
  \country{Germany}}
\email{michael.faerber@kit.edu}

\begin{abstract}
Environmental, social and governance (ESG) engagement of companies moved into the focus of public attention over recent years. With the requirements of compulsory reporting being implemented and investors incorporating sustainability in their investment decisions, the demand for a transparent and reliable ESG ratings is increasing. However, automatic approaches for forecasting ESG ratings have been quite scarce despite the increasing importance of the topic. In this paper 
we build a model to predict ESG ratings from news articles using the combination of multivariate timeseries construction and deep learning techniques. A news dataset for about 3,000 US companies together with their ratings is also created and released for training.
Through the experimental evaluation we find out that our approach provides accurate results outperforming the state-of-the-art, and can be used in practice to support a manual determination or analysis of ESG ratings.
\end{abstract}



\settopmatter{printfolios=true} %

\maketitle

\section{Introduction}

Companies' environmental, social and governance (ESG) practices are increasingly coming to the fore of the general public, institutional investors, and politics, while investments in ESG compliant or sustainable assets gain in popularity. For instance, in the US, assets invested in sustainable funds amounted to more than \$330 billion in 2021 \citep{cnnEvolutionESGinvest}. 
ESG and related topics are not only relevant to companies and investors but also to governments and society which increasingly focus on ESG compliance when deciding on contracts or making purchasing decisions. 
To better grasp the meaning and scope of the term ESG, we quote below its general definition provided by the International Finance Institute (IFC) \citep{IFC_Guidebook}:

\begin{it}
"A set of environmental, social and governance factors considered by companies when managing their operations, and investors when making investments, in respect of the risks, impacts and opportunities relating to but no limited to:
\begin{itemize}[leftmargin=.5cm]
    \item Environmental issues: potential or actual changes to the physical or natural environment (e.g. pollution, biodiversity impacts, carbon emissions, climate change, natural resource use)
    \item Social issues: potential or actual changes on surrounding community and workers (e.g. health and safety, supply chain, diversity and inclusion)
    \item Governance issues: corporate governance structures and processes by which companies are directed and controlled (e.g. board structure and diversity, ethical conduct, risk management, disclosure and transparency), including the governance of key environmental and social policies and procedures." \citep{IFC_Guidebook}
\end{itemize}
\end{it}

For many companies, improving their ESG performance and reducing ESG related risks has become an important part of long-term strategies \cite{zumente2021esg}. Especially negative ESG news and misconduct poses both reputational and financial risks to corporations. 
According to estimates of the Bank of America, S\&P500 companies lost more than \$600 billion of market value due to ESG issues over the last seven years \citep{BoFa_ESG_matters}. With respect to risk management, the importance of ESG increases from an reputational perspective, too. Even companies with a high value or reputation might not be able to bear consequences of negative ESG headlines as this potentially leads to brand damage, loss of customers and lower investments. These risks are further increasing as the awareness with regards to sustainability topics grows among investors, governments and other stakeholders. 

Predicting ESG ratings without including human judgment but fully automatically by leveraging NLP algorithms should decrease the cost for companies, making it -- above all -- also available for small or medium enterprises. Furthermore, automatic approaches should ensure that the ratings are transparent and can be potentially reconstructed by any stakeholder.

Despite the usefulness and importance of 
ESG analysis, the research on the assessment of firms' ESG performance is however quite scarce \cite{sokolov2021building,borms2021semi,esg1,esg2}. Sokolov et al. \cite{sokolov2021building} focus on the classification of Twitter content with respect to ESG relevance proposing simple approaches to construct indices based on classification results. 
While \citet{borms2021semi} predict ESG scores using news data, they do not leverage machine learning algorithms to do so but rather use simpler aggregation functions. Additionally, the dataset the authors use is small and their method requires human judgment. Other solutions either do not employ deep learning or use relatively simple models like tuned Feedforward neural network \cite{esg1,esg2}.

Thus to fill in this research gap, the objective of our work is to provide a framework for predicting ESG ratings automatically and more efficiently using multiple timeseries and deep learning technologies.
First, we collect ESG ratings for about 3,000 US American companies from a third-party provider and then we generate a corresponding news dataset of over 3,700,000 articles using the Global Database of Events, Language, and Tone (GDELT). In the first stage of our pipeline, all the news articles are classified with respect to their ESG relevance. The second step focuses on determining the articles' sentiment while the third step represents the articles with respect to their content. All these steps should provide broad information useful for assessing companies' ESG stance. In the fourth step, different models to predict ESG ratings therefrom are adopted based on the multi-variate timeseries constructed from the results of the prior steps. 

To sum up, our work makes the following contributions:

\begin{enumerate}[leftmargin=.5cm]
    \item We create ESG-related news dataset for about 3,000 US companies for the years 2018, 2019 and 2020. The dataset contains news articles from more than thousand different sources from various countries. To the best of our knowledge no comparable dataset in terms of size and diversity has been created and analyzed in the area of ESG related research. Studies relying on news datasets used mostly news provided by one specific outlet or published in a specific country \citep{sokolov2021weak, borms2021semi, nugent2021detecting}.
    \item We propose a multi-variate timeseries based approach that harnesses different signals from data based on sentiment analysis, semantic and topics analysis. We then use CNN and transformer-based models for generating the final ratings.
    \item We test our model against state-of-the-art, and perform thorough analysis of the model capability and characteristics.
\end{enumerate}

\section{Related Work}

The growing attention to ESG practices of companies and the trend of socially responsible investing in recent years arouse the interest of researchers in financial, economic and related areas (e.g., \cite{friede2015esg,gillan2021firms,kruger2015corporate,li2020difference,papoutsi2020does}).

Computational approaches to ESG rating analysis and prediction have been however quite scarce. Analyzing corporate social responsibility (CSR) reports, \citet{kiriu2020text} use text mining to evaluate companies' ESG activities. They focus on the quantity and specificity of ESG activities based on qualitative information provided in CSR reports. Using word2vec word embeddings, they classify the words extracted with respect to the three ESG dimensions, environmental, social and governance. 
The quality of ESG activities is measured by \cite{kiriu2020text} by assessing the divergence of words in the companies CSR reports. 

Instead of analyzing CSR reports, hence information provided by the company itself, \citet{azhar2019text} conduct a content analysis of news articles. They show that companies' CSR prominence can be assessed based on information in the news. Using news as data source mitigates issues like under- or overreporting of ESG / CSR topics according to them. 
\citet{hisano2020prediction} use news as information source too, arguing similar to \citet{azhar2019text} that these are less susceptible to manipulation. They utilize a network-based approach
with the goal to predict the appearance of a company on an investment exclusion list due to ESG issues. 
The approach of distant supervision of neural language models is pursued by \citet{raman2020mapping}. Using the transcripts of earnings calls, they evaluate which percentage of discussions in earnings calls focused on ESG topics over the last 5 years. 

When it comes to the actual task of ESG rating prediction, we are only aware of few works.  \citet{esg1} employes an ensemble approach over models like 
XGBoost, CatBoost and a Feedforward NN, while \citet{esg2} utilizes random forrest algorithm for ESG rating prediction.
\citet{borms2021semi} show that ESG indices constructed from a Flemish-Dutch news articles can predict negative adjustments in ESG scores 
up to a few months in advance. Applying a semi-supervised text mining approach they construct ESG indices for 291 European companies from a dutch-written news corpus containing 365,319 articles after dropping irrelevant articles. In the first step of their analysis, a set of keywords is generated for which the articles are queried in the next step to assess their ESG relevance. Defining the seed words for the three dimensions, environmental, social and governance, as well as the additional negative sentiment dimension is done manually making use of factors defined as important. 
This set of keywords is then enlarged leveraging GloVe word embeddings of their news corpus, prefiltered per dimension. The actual selection of keywords is again done partly manually. 
To transform the information contained in the news corpus into indices, \citet{borms2021semi} set up a matrix,
consisting of eleven frequency-based and 6 sentiment-adjusted indicators. 
Whereas \citet{borms2021semi} apply rather basic NLP techniques, by using GloVe word embeddings, Sokolov et al. \cite{sokolov2021building} leverage a %
the Bidirectional Encoder Representations from Transformers (BERT) to analyze social media data (tweets). The authors show that the accuracy of classifying ESG relevant content can be improved using a 
standard BERT classifier
with an additional fully connected layer before the output layer. 
For labeling the data, prior to model training, 
they prefiltered each of their ESG categories using keywords and human review. Thereby they select 1,468 ESG relevant tweets, labeled either positive or negative, out of the 6,000 tweets initially derived from Twitter for the time period from June 16th to July 22nd 2019. 
The authors propose an aggregation function for ESG index construction. The daily index of a company is derived by dividing the sum of each category's predicted probabilities by the sum of input documents for the day. Even though the authors of \cite{sokolov2021building} propose this method for ESG index construction, they do not report the resulting indices nor compare them to any benchmark. %

Our approach is based on the idea of predicting ESG ratings from news data like in \cite{borms2021semi}. However, 
compared to that method we utilize multi-variate timeseries solution that captures different aspects of companies including the changes in the number of relevant and irrelevant articles, sentiment and semantics of their ESG related information. The resulting timeseries are then used as input to train a deep learning model on the task of predicting ESG rating. 
Furthermore, our solution is a fully automated approach, which does not need any manual selection 
that is necessary in the methods of \cite{borms2021semi} and \cite{sokolov2021building}. Finally, we use at least a magnitude larger dataset, both in terms of the number of companies and number of news articles. Our results indicate that constructing timeseries based on different NLP analysis steps and feeding these as input to a deep learning model, is a promising model architecture for the ESG prediction task.

\vspace{-.5em}
\section{Dataset}
In this section we first describe the collection of ESG ratings for selected US companies and then the creation of news article dataset that corresponds to those ratings.
\vspace{-.5em}
\subsection{ESG Ratings Dataset Creation}\label{esgdata}

Asset4 ESG ratings\footnote{\url{https://www.refinitiv.com/en/sustainable-finance/esg-scores}} provided by Refinitiv, which is a global provider of financial market data and infrastructure, are used as reference point for the predictions. We have retrieved yearly ratings for all US companies rated by Refinitiv for 2018-2020, in total for 3,343 companies.  Only US American companies are considered to allow focusing on news in English language.

Three different ratings are provided by Refinitiv: (1) basic ESG scores, (2) ESG controversies scores, and (3) combined (controversy adjusted) ESG scores. All three ratings have values between 0 to 100 (the higher, the better the company's ESG performance). The basic ESG score mainly relies on information published in annual and CSR / ESG reports of the company. The creators of the scores defined 10 main themes grouped into the three dimensions of ESG that are measured by 186 specilized metrics:

\begin{it}Environmental:\end{it} resource use, emissions, innovation
 
\begin{it}Social:\end{it} workforce, human rights, community, product responsibility

\begin{it}Governance:\end{it} management, shareholders, CSR strategy 

The ESG controversies score is based on information from global media, and its value is the lower, the more negative and controversial news about the company are identified. Finally, the combined ESG score represents the basic score, adjusted by controversies providing a complete "picture" of company ESG activities. The lower the controversy score, the higher the downward adjustment of the combined score \citep{RefinitivMethodology}. We use the combined score as our groundtruth scores.

Additionally to the ratings itself, the long and short version of the company name for subsequent retrieval step as well as the market capitalization is derived from Refinitiv for each company. 
The market capitalization\footnote{Market capitalization is calculated by multiplying the price of a companies' stock by the total number of outstanding shares.}, referred to as company size in the financial industry is also recorded as it will be used later
to analyse the effect of company size on the precision of predictions.\footnote{We divide companies into 3 groups, small caps, mid caps and large caps based on their market capitalization. Even though no official thresholds exist, companies with a market capitalization greater than \$10 billion are considered as large caps, with values between \$2 to \$10 billion as mid caps and with values below \$2 billion as small caps \citep{MarketCap_def}.}

\subsection{News Dataset Creation}
To predict ESG ratings from news articles, a large dataset of news covering ESG-related topics is necessary. For this we utilize the GDELT project database \citep{GDELT} used by researchers across various disciplines \citep{azhar2019text, wu2017forecasting}.
The GDELT project provides a vast database covering a wide range of news (print, web, broadcast, television) from most countries around the world in more than 100 different languages. It is updated every 15 minutes. 
Using the GDELT Doc API Client, we have collected a list of the top matching 250 articles for
the time unit defined, over the three years: 2018, 2019 and 2020. 
The time unit was set to one month, hence up to the top 250 matches per month for the defined filter criteria were included in the dataset. Choosing 250 articles per month provided the most promising results.\footnote{Explorative analysis showed that shorter periods like days or weeks disproportionately increase the number of unrelated articles in the match list.} As a result, the initial dataset for each company contains at most 3,000 articles per year. 
The chosen keyword combination for collecting articles includes the 10 main themes of the three dimensions of ESG (see Sec. \ref{esgdata}) as well as a short and a long version of the company name (e.g., 'Visa Corporation' and 'Visa'). 
Similar to \cite{borms2021semi}, we required that the name of the company is mentioned at least two times in each matching article to exclude articles which refer to the company only in passing. 

Based on the list of URLs returned by querying GDELT, we next crawl the news articles.
We have then preprocessed the dataset removing articles shorter than 200 characters, articles with 'bad HTML requests' and extracting text content using Beautiful Soup\footnote{\url{https://pypi.org/project/beautifulsoup4/}} library. 
We have also removed companies for which less than 5 news articles were collected.
The final cleaning step removed further noise and unwanted characters, resulting from scrapping. All links, e-mail addresses, date and time representations as well as Unicode characters and new line representations were removed. 
Due to the length limitation of BERT models which we use, we utilized the title as well as the first 5 sentences from each news article.

\autoref{overview_news} provides an overview of the article dataset after cleaning. 
The majority of news articles (72\%) collected originate from the United States. News out of the United Kingdom, India and Canada represent 7\%, 3.4\% and 2.5\% of total articles, respectively.
Looking at the distribution among domains no major source could be identified, with the two largest domains, \begin{it}prnewswire.com\end{it} and \begin{it}news.yahoo.com\end{it}, accounting for 7\% and 4\% of total articles, respectively. 
After cleaning the data as described above, the final dataset comprises 3,739,871 articles for 8,484 ratings. 
\begin{table}[tb]
\centering
\caption{Overview of the News Article Dataset}
\label{overview_news}
\begin{small}
\begin{tabular}{lrrr}
& \textbf{2018} & \textbf{2019} & \textbf{2020}
\\ 
\hline
\textbf{\# of companies}            & 2,665                            & 2,940                            & 2,879                               \\
\textbf{\# of articles}             & 1,165,840                         & 1,399,277                         & 1,174,754                                  \\ 
\hline
\textbf{\# of articles per company} &                                   &                                   &                                                         \\
\textit{\textbf{average}}           & 435                               & 473                               & 406                                   \\
\textit{\textbf{max}}               & 1,984                             & 2,154                             & 2,005                                \\ 
\hline
\end{tabular}%
\end{small}
\end{table}

\section{Methodology}

\begin{figure}[tb]
 \centering
 \includegraphics[width=260pt]{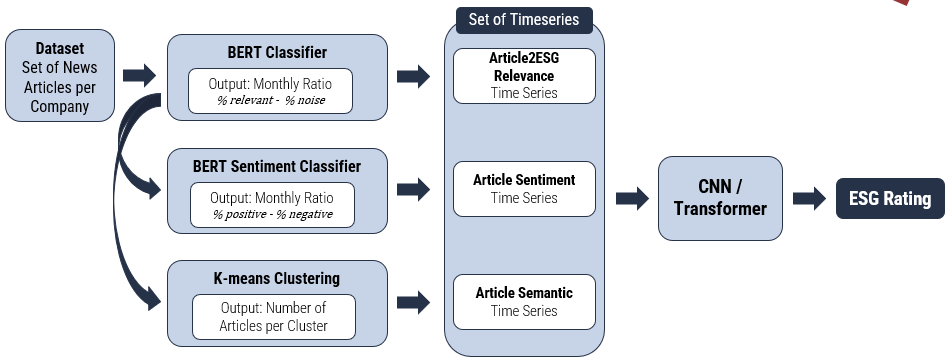}
  \caption{Overview ESG Prediction Model}
    \label{model_pic}
\end{figure}

An overview of our approach is provided in \autoref{model_pic}.
In the first step, articles are classified into ESG relevant and irrelevant articles. Deriving the sentiment, separately for relevant and irrelevant articles per company, is the second step. In third step, articles are clustered with respect to their semantic similarity. Thereby, nine timeseries are created for each company: one Article2ESG Relevance timeseries, two Article Sentiment timeseries as well as six semantic timeseries (one per each cluster). These timeseries are created separately for each company and each year. 
To predict the news-based ESG ratings, different neural network configurations, including the Transformer architecture and convolutional layers, are applied to model the input timeseries. Below we describe each of these steps.

\vspace{-8pt}
\subsection{Model Inputs: Text Classification}
To determine whether an article contains relevant information about companies' ESG performance and activities is a crucial step of the pipeline. 
We use BERT and fine-tune it for the specific task of ESG relevance classification. 
As labeling a few thousand articles by hand would be time-consuming we resorted to a weak-supervision approach.
This means a two step process to classify the articles with respect to their topical relevance. In the first step, a subset of the data is pre-labeled by determining the articles' semantic similarity to texts defining each of the three ESG dimensions. The ESG definitions of the Corporate Finance Institute\footnote{The CFI is a highly regarded institution in the financial profession providing training and certification programs among others. The definitions can be found on its webpage: \url{https://corporatefinanceinstitute.com/resources/knowledge/other/esg-environmental-social-governance/}} are utilized for the labelling. 

The news article subset, which is pre-labeled and used to train, validate and test the classifier in the next step, is composed of articles about 100 randomly sampled companies. Pre-trained text embeddings for the ESG definition sequences and the news articles are constructed exploiting the sentence transformer model 'all-distilroberta-v1' (SBERT) provided by Huggingface \citep{wolf2019huggingface, reimers2019sentence}. 
The labeling of the data as ESG relevant or irrelevant is then conducted by calculating the pairwise cosine similarity between an article's content and each of the 10 ESG category descriptions. 
All articles with a similarity value above 0.1 are labeled as relevant.\footnote{This rather low threshold is chosen due to two reasons. First, the ESG definitions are general, hence setting the threshold too high would be too selective. %
Second, we tested different threshold levels and found that the value of 0.1 provides a distribution of roughly 54\% relevant and 46\% noise while higher thresholds decrease the number of relevant articles significantly. For example setting 0.2 as threshold leads to a distribution of 17\% relevant to 83\% noise. A relatively large number of relevant articles seems
reasonable given the fact that the article set per company was already selected for the keywords relevant to environmental, social and governance issues when querying GDELT and therefore a certain topical relevance can be assumed.}

The second step of the process comprises fine-tuning a pre-trained classifier based on DistilBERT \cite{sanh2019distilbert}, and predicting the labels for the whole dataset using the resulting model. 
To tokenize the articles the DistilBertTokenizer provided by Huggingface is used \citep{wolf2019huggingface}. 
The tokenized inputs are then feed into the DistilBERT model to derive the embedding vectors using padding in case of too short articles. 
The last hidden state of the DistilBERT model corresponding to the classifier token 'CLS', is selected as input to the following layers.
The derived 2-dimensional vector is further processed by a dense layer of size 265 and a following dropout layer with dropout rate 0.5. 
The final dense layer of size 2 uses the softmax activation function. 
This model architecture is inspired by \cite{sokolov2021building}.\footnote{We compared it to other architectures by adding, for example, an additional dense layer or a bidirectional LSTM layer between the embedding and the dense layer, however these did not improve the results.} 

The DistilBERT classifier is fine-tuned in two steps. In the first step only the customized classification layers are optimized, while in the second step the whole network is fine-tuned using a lower learning rate \citep{sokolov2021building, sokolov2021weak}. The model parameters, including early stopping criteria are shown in Appendix. %
Both training steps are run for 4 epochs, however, the second stage is stopped after 2 epochs for most model configurations.
Using the trained classification model, the label of each article is then predicted. 

The Article2ESG Relevance timeseries, which represents the results of the classification analysis for detecting relevant and irrelevant articles, is calculated by first grouping the articles by month and company. %
The monthly value for Article2ESG Relevance is then calculated according to the following formula:
\begin{align}
  rel-noise_m = (Relevant_m / N_m) - (Noise_m / N_m)
\end{align}
The monthly relevance to noise ratio, $rel-noise$\textsubscript{$m$}, is 
represented by the subtraction of the monthly noise from the monthly relevance ratio. $Relevant$\textsubscript{$m$} is the number of relevant articles counted for the selected company in the considered month $m$. $Noise$\textsubscript{$m$} represents the number of noise articles respectively. $N$\textsubscript{$m$} is the total number of articles counted in month $m$. The difference instead of the ratio is taken to consider issues of the noise ratio being zero. 
Applying subtraction leads to a positive $rel-noise$\textsubscript{$m$} if the percentage of ESG relevant articles in month $m$ is larger than the percentage of noise articles and to a negative value otherwise. The values then range between -1 (only noise articles) to 1 (only relevant articles). As a result, a companies' Article2ESG Relevance Timeseries contains 12 $rel-noise$\textsubscript{$m$} values, one for each month of the considered year.

\subsection{Sentiment Analysis}
In a second step, the sentiment of the labeled ESG relevant articles is analysed. 
We use SieBERT model \citep{heitmann2020more} 
to predict the articles' sentiment. 
The labels returned by the model are used to generate the monthly Article Sentiment Timeseries per company. The sentiment timeseries is calculated separately for the ESG relevant and irrelevant articles. %
The articles of each corpus are grouped by company and month to calculate a monthly sentiment ratio: 
\begin{align}
  pos-neg_m = (Positive_m / N_m) - (Negative_m / N_m)
\end{align}
Similar to the $rel-noise$\textsubscript{$m$} ratio (Eq. (1)), the $pos-neg$\textsubscript{$m$} ratio is 
calculated by subtracting the negative ratio from the positive ratio. 
The Article Sentiment Timeseries are finally build by concatenating the monthly values for each company to a timeseries containing 12 values for the respective year.

\subsection{Semantic Analysis}

The third step, the semantic analysis, aims at grouping the articles with respect to their content in order to identify the evolution of topics discussed in the news. 

First, the DistilBERT algorithm is fine-tuned on the task of ESG classification. %
The resulting task-specific embeddings of articles are then used as input to k-means clustering algorithm.\footnote{Additionally to 
k-means algorithm, we also tested the spherical k-means algorithm but the results did not change much.} 
To determine the number of clusters $k$ we apply the elbow method which determined 6 as the number of appropriate clusters.

We include the results of the semantic analysis for the input to the final prediction of the ESG ratings, as timeseries, each one for different cluster. Hence in total 6 semantic timeseries are created per company according to the following steps. First, the dataset is grouped by clusters to derive the total set of articles belonging to each cluster. In the next step, each cluster subset is grouped by company to derive the overall size of each cluster per company. The monthly values of the timeseries of cluster $l$ are calculated by counting the number of articles per month belonging to the respective cluster $l$. 
The 12 monthly values per company and cluster are concatenated, to derive 6 semantic timeseries per each company. Thereby the monthly size of each 
cluster identified is tracked, representative for the main topics identified and how dominant these are for a respective company.

\subsection{Rating Prediction}
In this section we discuss 4 models proposed to generate ESG rating based on the prepared timeseries.

\begin{figure}[t]
 \centering
   \includegraphics[width=205pt]{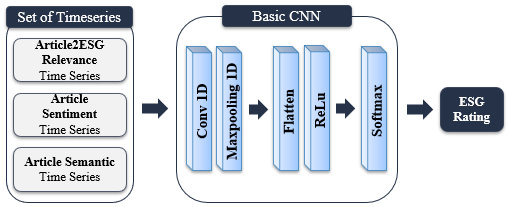}
    \caption{Basic CNN Model}
     \label{cnn_v1}
\end{figure}

The first and the simplest model that we experiment with is a convolutional neural network shown in Fig. \ref{cnn_v1}.
The nine constructed input timeseries for 12 months that were discussed before: the Article2ESG relevance, the Article Sentiment (2 timeseries) as well as the Article Semantic timeseries (one per each of 6 clusters), constitute the input to the model in form of a matrix (9 x 12 dimensions). Before being fed into the model, the timeseries are standardized using standard scaling (i.e., substracting the mean and diving by the standard deviation). The standardized timeseries are then processed by a convolutional 1D layer, followed by a maxpooling 1D layer. 
The resulting output of these two layers is then flattened and
a dense layer with a ReLu activation is added to introduce non-linearity to the model \citep{chollet2021deep}.
In the final classification layer, the softmax function is applied to return the probability of each of the $n$ classes (i.e., ESG rating values). For each sample, the class with the maximum probability is selected as the final ESG rating. 

\begin{figure}[tb]
 \centering
   \includegraphics[width=250pt]{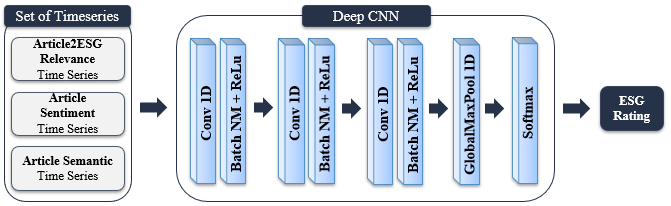}
    \caption{Deep CNN Model}
     \label{cnn_v2}
\end{figure}

Additionally to this simple convolutional neural network comprising only a single convolutional layer, a deeper CNN is also introduced. It includes 3 convolutional blocks as illustrated in \autoref{cnn_v2}. 

These three blocks are built of three layers each, a convolutional layer, a batch normalization layer and a ReLu activation. 
Adding a batch normalization layer serves to standardize the layers' input, and several comparisons show that this helps to speed up the models training \citep{chollet2021deep}. The three convolutional blocks are followed by a global max pooling layer. %
The hyperparameters, which yielded the best results for the two CNN models, are displayed in the Appendix. 

\begin{figure}[tb]
 \centering
   \includegraphics[width=240pt]{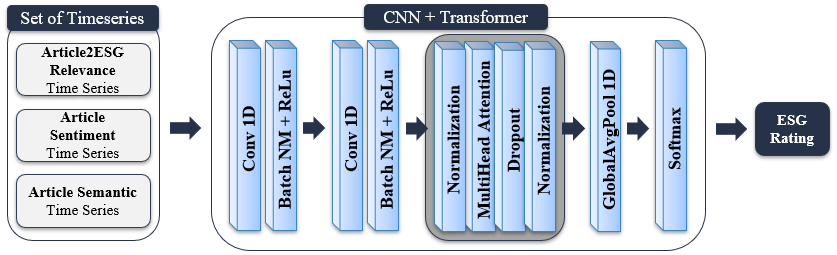}
    \caption{CNN with a single Transformer Layer}
     \label{TF}
\end{figure}

Besides the convolutional neural networks, 
we use also two other models, a basic model including CNN and a single Transformer layer and a deeper model with CNN and a multi-layer Transformer encoder block. The basic model, which includes two convolutional blocks as well as a single Transformer layer, is shown in \autoref{TF}. 

 The transformer encoder block \cite{vaswani2017attention} consists of a multi-head attention, a dropout and two normalization layers (highlighted in gray in \autoref{TF}). 
 Normalization is applied to the input, which refers to the output of the second convolutional block, as well as to the output of the attention block. 
 The central layer is the multi-head attention layer with added dropout layer.
 Like for the deep CNN, the output is transformed into class predictions by applying global maxpooling and the softmax function. The deeper, multi-layer transformer model, has a similar architecture as shown in \autoref{TF} with the difference that the single Transformer block is replaced by three consecutive transformer encoder blocks. 

The model parameters for both the basic and the deep Transformer model are presented in Appendix.
The CNN part comprises two convolutional layers of filter size 64 and 128 with kernel size 3 and 1, respectively. 
The same parameters are used for all Transformer layers, they consist of 8 heads of size 200 each and a dropout of 0.2 is applied. 

The four model architectures shown can be referred to as timeseries classification models, where each possible rating refers to one class. In the case of the Asset4 ratings used for comparison, ratings range from 0 to 100, hence they can be treated as 100 classes. 
Nonetheless, the ESG rating prediction can be considered as a regression problem, too. 
We test both the approaches. For regression, the last layer, the softmax layer, is replaced by a linear dense layer, which only returns one output, being the predicted rating as a continuous number between 0 to 100.

\section{Experiments}
After first discussing experimental settings, we will next analyze the different components of the multi-variate timeseries input (Sec. 5.2 - 5.4) and then we focus on ESG rating prediction's accuracy (Sec. 5.5-5.7).
\subsection{Experimental Settings}
We compare our approach with several baselines including random selection, majority class as well as the method proposed by \cite{sokolov2021building} applied on our dataset. As mentioned before, \cite{sokolov2021building} use the output of their ESG classification model, and a BERT classifier, to construct their ESG scores. More specifically the scores are calculated by taking the average of the predicted probabilities for each set of input documents per company and day. 
Since the Asset4 ratings are on a yearly basis, the daily scores will be aggregated to a yearly score for comparison. This is done by taking the average over the daily predictions per year.

Common model parameters, which are applied to all the four proposed models in order to ensure a certain level of comparability, are listed in Appendix. %
The rectified Adam optimizer proposed by \cite{liu2019variance} is used, which provides improved results compared to earlier emerged optimizers. 
The standard learning rate of 1e-3 is applied. To mitigate the issue of overfitting, early stopping criteria are applied. If five consecutive epochs do not improve the validation loss, training is stopped. Additionally the learning rate is reduced if five consecutive steps show only a minor change of the validation loss below the threshold of 0.01. Generally, all models are trained for a maximum of 25 epochs, however, early stopping criteria are triggered earlier for most models. The Transformer regression model with three Transformer blocks makes an exception, it is trained for a maximum of 50 epochs as training continues and results improve thereby. 

For the classification models, sparse categorical accuracy and sparse categorical cross entropy loss are used.  
Mean average percentage error and mean squared error are used by the regression models, as these are commonly applied to evaluate regression performance. %
Whereas the returned metrics and losses by the classification and regression models are not directly comparable, the deviation from the Asset4 ratings is directly comparable. As the differences between the Asset4 and the predicted ratings can be positive or negative, the average of the absolute differences is taken for comparison.

\subsection{ESG News classification}

We first look into the classification accuracy of ESG-related articles that was used for building the Article2ESG relevance timeseries.
\autoref{distilBERT_test} reports the training statistics of the DistilBERT model with the chosen configuration. Different model parameters were tested, the best results are retrieved with using the maximum input length 512 of the model, a batch size of 32, a learning rate of 1e\textsuperscript{-4} for the first training stage and 2e\textsuperscript{-5} for the second training stage. The model is trained for 4 epochs per stage, however, the second stage shows that the model starts to overfit after the second epoch. 
The model was trained with about 10,500 articles and a validation set of size 2,100.
Using the fine-tuned weights, the build DistilBERT classification model is evaluated based on a test set of 57,088 articles. 
This represents the remaining number of articles in the subset of articles, from 100 random sampled companies, after removing train and validation data. 
The classification report shown in \autoref{distilBERT_test} reports an overall accuracy of 83\%. Focusing on the class of interest, the ESG relevant articles, a precision of 85\% is reached at a recall of 86\%. The prediction of the noise or irrelevant class shows a precision of 80\% at a recall of 79\%. Although not perfect, these values should be good enough for generating the relevance timeseries to be incorporated into the input of the model.

\begin{table}[tb]
\centering
\caption{Classification Report}
\label{distilBERT_test}
\resizebox{0.4\textwidth}{!}{%
\begin{tabular}{rcccc}
\hline
\textit{Max Length}           & \multicolumn{1}{c}{\textit{512}}       & \multicolumn{1}{r}{\textit{Training Epochs}} & \multicolumn{1}{c}{\textit{4}}        & \multicolumn{1}{c}{\textit{2}}       \\
\textit{Batch Size}           & \multicolumn{1}{c}{\textit{32}}        & \multicolumn{1}{r}{\textit{Learning Rate}}   & \multicolumn{1}{c}{\textit{1e-4}}     & \multicolumn{1}{c}{\textit{2e-5}}    \\ \hline
\multicolumn{1}{c}{\textbf{}} & \multicolumn{1}{c}{\textbf{precision}} & \multicolumn{1}{c}{\textbf{recall}}          & \multicolumn{1}{c}{\textbf{f1-score}} & \multicolumn{1}{c}{\textbf{support}} \\ \hline
\textbf{noise}                & 0.80                                   & 0.79                                         & 0.80                                  & 23,938                               \\
\textbf{ESG relevant}         & 0.85                                   & 0.86                                         & 0.85                                  & 33,150                               \\ \hline
\textbf{accuracy}             &                                        &                                              & 0.83                                  & 57,088                               \\
\textbf{macro avg}            & 0.83                                   & 0.83                                         & 0.83                                  & 57,088                               \\
\textbf{weighted avg}         & 0.83                                   & 0.83                                         & 0.83                                  & 57,088                               \\ \hline
\end{tabular}%
}
\end{table}

\subsection{Sentiment Analysis}

\begin{figure}[tb]
  \centering
    \includegraphics[width=200pt]{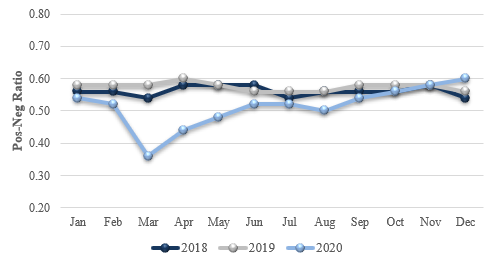}
    \caption{Overview of pos-neg ratios by month}
\label{pos_neg_m}
\end{figure}

Assessing the overall sentiment of the articles represents the second level of analysis. 
An overview over the average monthly pos-neg values for the subset of ESG relevant articles is provided in \autoref{pos_neg_m}.
The monthly pos-neg values range between 0.5 to 0.6 for most months, indicating that the positive overweight the negative article significantly among the ESG relevant data subset. The distribution is similar for the noise subset of articles. About three times more articles are assigned a positive compared to a negative sentiment label (we show the detailed distributions in Appendix).
The large drop in the ratio reported for March 2020 and the lower values for April and May, are related to Covid-19. These months were characterized by the beginning of the Covid-19 pandemic and the resulting severe restrictions, which had a negative impact on companies and their operations. The basic tone of news was more negative in these months.

\subsection{Cluster Analysis}
The third level of analysis builds the semantic analysis of the input texts. The article contents in this case are used to cluster the articles into six groups. 
The clusters are identified for each year separately.

\begin{figure}[tb]
  \centering
    \includegraphics[width=170pt]{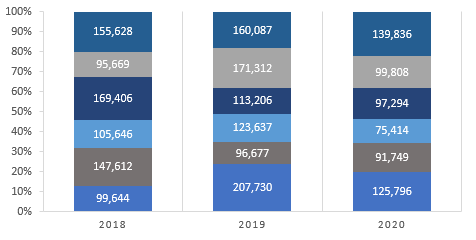}
    \caption{Overview of cluster sizes
    }
\label{cluster_yearly}
\end{figure}

The size of the created yearly clusters is shown in \autoref{cluster_yearly}. The article distribution is spread rather evenly across each of the 6 clusters. This applies to 2018, 2019 and 2020. It needs to be kept in mind that the clusters are created separately for each year, hence cluster 1 in year 2020 is not comparable to cluster 1 in year 2021. No dominant cluster can be identified for any of the years analysed, rather between 12\% to 22\% of the data is mapped to each of the 6 clusters. 

Further we analyzed the composition of the clusters in more detail. For each cluster its articles, are grouped by company to derive the number of articles per company included in each cluster.´We find that no single company is dominant in any cluster; the companies with the largest share only represent between 0.8\% to 1.5\% of the respective cluster and all clusters contain articles from more than 2,000 different companies. Nevertheless, focusing on the companies which contribute the largest proportion to a selected cluster, reveals that the companies seem to be related to each other by their industry. For example, in the second cluster in 2018 \textit{Via Renewables, Clean Energy Fuels, Duke Energy, Concophillips} and \textit{Advanced Energy} are among the top 10 companies, which all belong to the energy sector. 
It emphasizes that the articles' semantics are industry specific to some extent.

\subsection{Evaluation of Predicted ESG Rating}

As described in detail in Section 4.4, four different models are applied to predict ESG ratings from the constructed timeseries, in two different configurations, as classification and as regression models. \autoref{eval_2020} shows the results for all model variations as well as baselines models for the year 2020. The reported results represent the average values over 10 runs, conducted with different, randomly initialized train-test splits. Because evaluation metrics for classification and regression models are not directly comparable, an additional evaluation metric is considered. More precisely, the average absolute difference between the Asset4 ratings and the predicted ratings is calculated to compare the models based thereon. For the baseline models all model metrics except for the sparse categorical cross-entropy loss (scc) are calculated and reported. The scc cannot be calculated as class probabilities would be needed therefore.
Looking at the classification models, the accuracy values are quite low. However, the task to be completed, predicting the right out of 100 classes is quite complex and results vary significantly depending on the train-test split.

\begin{table}[tb]
\centering
\caption{Model Results for the year 2020}
\label{eval_2020}
\resizebox{0.45\textwidth}{!}{%
\begin{tabular}{rrrrrrrrrr}
\cline{3-6} \cline{8-10}
                                                            &                                        & \multicolumn{4}{c}{\textit{\textbf{Overall Results}}}                                                                                        & \multicolumn{1}{l}{\textit{\textbf{}}} & \multicolumn{3}{c}{\textit{\textbf{\begin{tabular}[c]{@{}c@{}}Absolute Difference \\ to Asset4 Rating\end{tabular}}}} \\ \cline{1-1} \cline{3-6} \cline{8-10} 
\multicolumn{1}{l}{\textit{\textbf{Classification Models}}} & \multicolumn{1}{l}{\textit{\textbf{}}} & \multicolumn{1}{c}{\textbf{acc}} & \multicolumn{1}{c}{\textbf{scc}} & \multicolumn{1}{c}{\textbf{}}     & \multicolumn{1}{l}{}             & \multicolumn{1}{l}{}                   & \multicolumn{1}{c}{\textbf{mean}}      & \multicolumn{1}{c}{\textbf{std}}      & \multicolumn{1}{c}{\textbf{max}}     \\ \cline{1-1} \cline{3-6} \cline{8-10} 
Basic CNN                                                   &                                        & 2.2\%                            & 4.29                             &                                   &                                  &                                        & 15.27                                  & 12.23                                 & 64.60                                \\
Deep CNN                                                    &                                        & \textbf{2.3\%}                            & \textbf{4.38}                            &                                   &                                  &                                        & 15.72                                  & 12.43                                 & 58.10                                \\ \cline{1-1} \cline{3-6} \cline{8-10} 
CNN + Transformer                                           &                                        & \textbf{2.3\%}                            & 4.49                             &                                   &                                  &                                        & 15.81                                  & 12.57                                 & 62.40                                \\
CNN + Deep Transformer                                            &                                        & 2.2\%                            & 5.29                             &                                   &                                  &                                        & 16.44                                  & 12.80                                 & 67.80                                \\ \cline{1-1} \cline{3-6} \cline{8-10} 
\multicolumn{1}{l}{\textit{\textbf{Regression Models}}}     & \multicolumn{1}{l}{\textit{\textbf{}}} & \multicolumn{1}{c}{\textbf{}}    & \multicolumn{1}{c}{\textbf{}}    & \multicolumn{1}{c}{\textbf{mape}} & \multicolumn{1}{c}{\textbf{mse}} & \multicolumn{1}{c}{}                   & \multicolumn{1}{c}{}                   & \multicolumn{1}{l}{}                  & \multicolumn{1}{l}{}                 \\ \cline{1-1} \cline{3-6} \cline{8-10} 
Basic CNN                                                   &                                        &                                  &                                  & 51.31                             & 285.07                           &                                        & 13.43                                  & 10.20                                 & 52.62                                \\
Deep CNN                                                    &                                        &                                  &                                  & 50.12                             & \textbf{263.18}                         &                                        & \textbf{13.04}                                  & 9.66                                  & 48.63                                \\
CNN + Transformer                                            &                                        &                                  &                                  & 44.18                             & 368.73                           &                                        & 15.05                                  & 11.88                                 & 54.41                                \\
CNN + Deep Transformer                                            &                                        &                                  &                                  & 50.66                             & 294.75                           &                                        & 13.36                                  & 10.59                                 & 73.38                                \\ \cline{1-1} \cline{3-6} \cline{8-10} 
\multicolumn{1}{l}{\textit{\textbf{Baselines}}}             & \multicolumn{1}{l}{\textit{\textbf{}}} & \multicolumn{1}{c}{\textbf{acc}} & \multicolumn{1}{c}{\textbf{scc}} & \multicolumn{1}{c}{\textbf{mape}} & \multicolumn{1}{c}{\textbf{mse}} & \multicolumn{1}{c}{}                   & \multicolumn{1}{c}{}                   & \multicolumn{1}{l}{}                  & \multicolumn{1}{l}{}                 \\ \cline{1-1} \cline{3-6} \cline{8-10} 
Mean Asset4 Rating                                          &                                        & 2.2\%                            &                                  & \textbf{43.64}                            & 365.63                           &                                        & 14.79                                  & 12.02                                 & 55.94                                \\
Random Selection                                            &                                        & 0.9\%                            &                                  & 102.79                            & 1167.99                          &                                        & 28.10                                  & 19.48                                 & 83.80                                \\
Sokolov et al. \cite{sokolov2021building}                                                     &                                        & 0.5\%                            &                                  & 176.27                            & 2108.27                          &                                        & 41.46                                  & 19.74                                 & 89.89                                \\ \cline{1-1} \cline{3-6} \cline{8-10} 
\end{tabular}%
}
\end{table}

When comparing all models based on the mean absolute difference, the best performing model is the proposed deep CNN regression model including three convolutional blocks. This model outperforms all others with respect to a lower mean absolute difference, a lower standard and low maximum deviation. With respect to the model metrics, it shows also the lowest mean squared error (mse) among the regression models. Only the mean average percentage error is higher. 
Furthermore, it is notable that three out of four regression models outperform all other models significantly based on the mean difference metric, with values of 13.04 and 13.34 compared to values from 15.05 up to 16.44. All regression models perform better than all classification models, which indicates that the prediction should be framed as a regression task. This is reasonable due to the fact that predicting a values slightly above or below the actual value is treated as wrong from a classification point of view while being treated as relatively good prediction from an regression point of view. 
We also note quite poor performance of the approach by Sokolov et al. \cite{sokolov2021building}.

The prediction results for the years 2018 and 2019 shown in \autoref{eval_2018} and \autoref{eval_2019}, respectively, paint a similar picture. Also here, we notice that the classification models perform worse than the regression models in 2018 and 2019, too. The deep CNN regression model is the best performing model over the analysed periods, outperforming all other models in the remaining two years. Generally, looking at the mean absolute difference to the Asset4 ESG ratings, the proposed models perform slightly better in 2018 and 2019, with mean difference values ranging from 11.85 to 15.91 compared to values between 13.04 to 16.44 in 2020.
The higher performance in 2018 and 2019 seems to be related to 
the Covid-19's interference with global economy in 2020. 

Additionally to the best performing deep CNN, the basic CNN and the deep Transformer regression model, which show promising performance in 2020 too, the basic CNN classification model outperforms all baselines in 2018 and 2019. The deep CNN classification model as well as the basic Transformer regression model perform better than all three baselines in the year 2018 only. The mean asset ratings represent the best baseline model over all three years, performing significantly better than random selection and the approach according to \cite{sokolov2021building}.

\begin{table}[H]
\centering
\caption{Model Results for the year 2018}
\label{eval_2018}
\resizebox{0.45\textwidth}{!}{%
\begin{tabular}{rrrrrrrrrr}
\cline{3-6} \cline{8-10}
                                                            &                                        & \multicolumn{4}{c}{\textit{\textbf{Overall Results}}}                                                                                              & \multicolumn{1}{c}{\textit{\textbf{}}} & \multicolumn{3}{c}{\textit{\textbf{\begin{tabular}[c]{@{}c@{}}Absolute Difference\\ to Asset4 Rating\end{tabular}}}} \\ \cline{1-1} \cline{3-6} \cline{8-10} 
\multicolumn{1}{l}{\textit{\textbf{Classification Models}}} & \multicolumn{1}{l}{\textit{\textbf{}}} & \multicolumn{1}{c}{\textbf{acc}}       & \multicolumn{1}{c}{\textbf{scc}} & \multicolumn{1}{c}{}              & \multicolumn{1}{c}{}             & \multicolumn{1}{c}{}                   & \multicolumn{1}{c}{\textbf{mean}}      & \multicolumn{1}{c}{\textbf{std}}     & \multicolumn{1}{c}{\textbf{max}}     \\ \cline{1-1} \cline{3-6} \cline{8-10} 
Basic CNN                                                   &                                        & \textbf{3.1\%}                                  & \textbf{4.23}                            &                                   &                                  &                                        & 14.35                                  & 12.66                                & 62.56                                \\
Deep CNN                                                    &                                        & 2.8\%                                  & 4.38                             &                                   &                                  &                                        & 14.55                                  & 12.79                                & 65.60                                \\ \cline{1-1} \cline{3-6} \cline{8-10} 
CNN + Transformer                                          &                                        & 2.8\%                                  & 4.49                             &                                   &                                  &                                        & 14.79                                  & 12.28                                & 65.30                                \\
CNN + Deep Transformer                                           &                                        & 2.6\%                                  & 5.59                             & \multicolumn{1}{l}{}              & \multicolumn{1}{l}{}             & \multicolumn{1}{l}{}                   & 15.91                                  & 13.18                                & 64.80                                \\ \cline{1-1} \cline{3-6} \cline{8-10} 
\multicolumn{1}{l}{\textit{\textbf{Regression Models}}}     & \multicolumn{1}{l}{\textit{\textbf{}}} & \multicolumn{1}{l}{\textit{\textbf{}}} & \multicolumn{1}{c}{\textbf{}}    & \multicolumn{1}{c}{\textbf{mape}} & \multicolumn{1}{c}{\textbf{mse}} & \multicolumn{1}{c}{\textbf{}}          & \multicolumn{1}{l}{}                   & \multicolumn{1}{l}{}                 & \multicolumn{1}{l}{}                 \\ \cline{1-1} \cline{3-6} \cline{8-10} 
Basic CNN                                                   &                                        &                                        &                                  & 49.75                             & 254.29                           &                                        & 12.44                                  & 9.94                                 & 59.77                                \\
Deep CNN                                                    &                                        &                                        &                                  & 47.35                             & \textbf{227.42}                           &                                        & \textbf{11.85}                                  & 9.27                                 & 53.07                                \\ \cline{1-1} \cline{3-6} \cline{8-10} 
CNN + Transformer                                           &                                        &                                        &                                  & \textbf{42.47}                             & 322.88                           &                                        & 13.67                                  & 11.62                                & 64.88                                \\
CNN + Deep Transformer                                             &                                        &                                        &                                  & 49.66                             & 243.50                           &                                        & 12.44                                  & 9.41                                 & 52.77                                \\ \cline{1-1} \cline{3-6} \cline{8-10} 
\multicolumn{1}{l}{\textit{\textbf{Baselines}}}             & \multicolumn{1}{l}{\textit{\textbf{}}} & \multicolumn{1}{c}{\textbf{acc}}       & \multicolumn{1}{c}{\textbf{scc}} & \multicolumn{1}{c}{\textbf{mape}} & \multicolumn{1}{c}{\textbf{mse}} & \multicolumn{1}{c}{\textbf{}}          & \multicolumn{1}{l}{}                   & \multicolumn{1}{l}{}                 & \multicolumn{1}{l}{}                 \\ \cline{1-1} \cline{3-6} \cline{8-10} 
Mean Asset4 Rating                                          &                                        & 2.0\%                                  & -                                & 62.82                             & 319.09                           &                                        & 14.62                                  & 10.25                                & 50.89                                \\
Random Selection                                            &                                        & 1.2\%                                  & -                                & 127.53                            & 1426.66                          &                                        & 31.07                                  & 21.48                                & 85.11                                \\
Sokolov et al. \cite{sokolov2021building}                                                     &                                        & 1.2\%                                  & -                                & 160.89                            & 1530.52                          &                                        & 33.62                                  & 20.01                                & 89.64                                \\ \cline{1-1} \cline{3-6} \cline{8-10} 
\end{tabular}%
}
\end{table}

\begin{table}[H]
\centering
\caption{Model Results for the year 2019}
\label{eval_2019}
\resizebox{0.45\textwidth}{!}{%
\begin{tabular}{rrrrrrrrrr}
\cline{3-6} \cline{8-10}
                                                            &                                        & \multicolumn{4}{c}{\textit{\textbf{Overall Results}}}                                                                                              & \multicolumn{1}{c}{\textit{\textbf{}}} & \multicolumn{3}{c}{\textit{\textbf{\begin{tabular}[c]{@{}c@{}}Absolute Difference \\ to Asset4 Rating\end{tabular}}}} \\ \hline
\multicolumn{1}{l}{\textit{\textbf{Classification Models}}} & \multicolumn{1}{l}{\textit{\textbf{}}} & \multicolumn{1}{c}{\textbf{acc}}       & \multicolumn{1}{c}{\textbf{scc}} & \multicolumn{1}{c}{}              & \multicolumn{1}{c}{}             & \multicolumn{1}{c}{\textbf{}}          & \multicolumn{1}{c}{\textbf{mean}}      & \multicolumn{1}{c}{\textbf{std}}      & \multicolumn{1}{c}{\textbf{max}}     \\ \hline
Basic CNN                                                   &                                        & 2.1\%                                  & \textbf{4.21}                             &                                   &                                  &                                        & 13.60                                  & 10.92                                 & 55.50                                \\
Deep CNN                                                    &                                        & 2.4\%                                  & 4.33                             &                                   &                                  &                                        & 14.39                                  & 11.86                                 & 59.50                                \\ \cline{1-1} \cline{3-6} \cline{8-10} 
CNN + Transformer                                           &                                        & \textbf{2.5\%}                                  & 4.44                             &                                   &                                  &                                        & 14.90                                  & 12.08                                 & 62.40                                \\
CNN + Deep Transformer                                            &                                        & \textbf{2.5\%}                                  & 5.23                             &                                   &                                  &                                        & 15.69                                  & 12.65                                 & 67.00                                \\ \cline{1-1} \cline{3-6} \cline{8-10} 
\multicolumn{1}{l}{\textit{\textbf{Regression Models}}}     & \multicolumn{1}{l}{\textit{\textbf{}}} & \multicolumn{1}{c}{\textit{\textbf{}}} & \multicolumn{1}{c}{\textbf{}}    & \multicolumn{1}{c}{\textbf{mape}} & \multicolumn{1}{c}{\textbf{mse}} & \multicolumn{1}{c}{}                   & \multicolumn{1}{l}{}                   & \multicolumn{1}{l}{}                  & \multicolumn{1}{l}{}                 \\ \cline{1-1} \cline{3-6} \cline{8-10} 
Basic CNN                                                   &                                        &                                        &                                  & 46.47                             & 242.51                           &                                        & 12.28                                  & 9.56                                  & 53.38                                \\
Deep CNN                                                    &                                        &                                        &                                  & 45.71                             & \textbf{227.83}                           &                                        & \textbf{11.97}                                  & 9.19                                  & 46.33                                \\ \cline{1-1} \cline{3-6} \cline{8-10} 
CNN + Transformer                                           &                                        &                                        &                                  & \textbf{40.25}                             & 348.70                           &                                        & 14.47                                  & 11.77                                 & 60.15                                \\
CNN + Deep Transformer                                          &                                        &                                        &                                  & 45.90                             & 245.45                           &                                        & 12.36                                  & 9.59                                  & 56.95                                \\ \cline{1-1} \cline{3-6} \cline{8-10} 
\multicolumn{1}{l}{\textit{\textbf{Baselines}}}             & \multicolumn{1}{l}{\textit{\textbf{}}} & \multicolumn{1}{c}{\textbf{acc}}       & \multicolumn{1}{c}{\textbf{scc}} & \multicolumn{1}{c}{\textbf{mape}} & \multicolumn{1}{c}{\textbf{mse}} & \multicolumn{1}{l}{}                   & \multicolumn{1}{l}{}                   & \multicolumn{1}{l}{}                  & \multicolumn{1}{l}{}                 \\ \cline{1-1} \cline{3-6} \cline{8-10} 
Mean Asset4 Rating                                          &                                        & \textbf{2.5\%}                                  & -                                & 55.88                             & 313.37                           &                                        & 14.39                                  & 10.40                                 & 49.70                                \\
Random Selection                                            &                                        & 0.9\%                                  & -                                & 120.66                            & 1436.93                          &                                        & 31.11                                  & 21.30                                 & 86.60                                \\
Sokolov et al. \cite{sokolov2021building}                                                     &                                        & 0.7\%                                  & -                                & 158.60                            & 1713.39                          &                                        & 36.34                                  & 20.49                                 & 90.93                                \\ \cline{1-1} \cline{3-6} \cline{8-10} 
\end{tabular}%
}
\end{table}
\subsection{Evaluation by Market Capitalization}

To provide further insights, 
\autoref{eval_19_cap_reg} presents the results for the regression models grouped by market capitalization (cf. Sec 3.1) for 2019. The 2019 results are representative for all three years, as the same patterns can be observed in 2018, 2019 and 2020.

Considering the mean absolute difference between the Asset4 ESG and the predicted ratings shown in \autoref{eval_19_cap_reg} we notice that the models make significantly better predictions for smaller compared to larger companies. For the best performing model, the deep CNN, the average deviation for small cap companies amounts to 10.28 compared to 15.42 for large caps.

\begin{table}[tb]
\centering
\caption{Absolute Difference Results for Regression Models by Market Capitalization (2019)}
\label{eval_19_cap_reg}
\resizebox{.4\textwidth}{!}{%
\begin{tabular}{llll}
\hline
\textbf{Regression Models} & \textbf{small cap} & \textbf{mid cap} & \textbf{large cap} \\ \hline
\textit{Basic CNN}        & 10.87                                  & 12.81                                & 15.25            \\ 
\textit{Deep CNN}         & 10.28                                  & 12.67                                & 15.42           \\ \hline
\textit{CNN + Transformer } & 12.16                                  & 14.64                                & 20.58           \\
\textit{CNN + Deep Transformer}  & 10.86                                  & 12.51                                & 16.09           
\end{tabular}%
}
\end{table}

\subsection{Ablation Study}

Additionally an ablation study is conducted in order to analyze the impact of the different timeseries and the related NLP analysis steps on the ESG rating prediction. \autoref{ablation} presents the results for the deep CNN regression model with varying inputs. It compares training the model using all nine input timeseries to using only the Article2ESG Relevance timeseries, only the Article Sentiment timeseries or using the six Article Semantic timeseries.

The best results are reported for X\textsubscript{all} as input, which represents the combination of all nine input timeseries. This indicates that each of the three analysis steps and the thereby generated timeseries improves the performance of the ESG rating prediction. Focusing on the three features, X\textsubscript{semantic} seems to be the most important feature, as it provides the best results when used as model input separately. However, we need to keep in mind that the X\textsubscript{semantic} feature includes all six cluster timeseries, whereas X\textsubscript{relevance} and X\textsubscript{sentiment} are a single and double timeseries, respectively. Hence, the effect could be also influenced by 
the larger size of the model input. 

\begin{table}[tb]
\centering
\caption{Results Input Ablation Study}
\label{ablation}
\resizebox{0.37\textwidth}{!}{%
\begin{tabular}{lrllllll}
\multicolumn{1}{l}{\textit{\textbf{}}}         & \multicolumn{1}{l}{\textit{\textbf{}}} & \multicolumn{2}{c}{\textit{\textbf{Model Metrics}}}                  & \multicolumn{1}{c}{\textit{\textbf{}}} & \multicolumn{3}{c}{\textit{\textbf{\begin{tabular}[c]{@{}c@{}}Absolute Difference\\ to Asset4 Rating\end{tabular}}}} \\ \hline
\multicolumn{1}{l}{\textit{\textbf{Features}}} & \multicolumn{1}{l}{\textit{\textbf{}}} & \multicolumn{1}{c}{\textbf{mape}} & \multicolumn{1}{c}{\textbf{mse}} & \multicolumn{1}{c}{\textbf{}}          & \multicolumn{1}{c}{\textbf{mean}}      & \multicolumn{1}{c}{\textbf{std}}     & \multicolumn{1}{c}{\textbf{max}}     \\ \hline
\textit{X\textsubscript{relevance}}                           & \textit{}                              & 53.01                             & 274.36                           &                                        & 13.29                                  & 9.90                                 & 50.03                                \\
\textit{X\textsubscript{sentiment}}                           & \textit{}                              & 51.81                             & 289.23                           &                                        & 13.55                                  & 10.28                                & 48.99                                \\
\textit{X\textsubscript{semantic}}                            & \textit{}                              & 47.97                             & 239.63                           &                                        & 12.22                                  & 9.51                                 & 51.84                                \\ \hline
\textit{\textbf{X\textsubscript{all}}}                            & \textit{\textbf{}}                     & \textbf{45.71}                    & \textbf{227.83}                  & \textbf{}                              & \textbf{11.97}                         & \textbf{9.19}                        & \textbf{46.33}                       \\ \hline
\end{tabular}%
}
\end{table}

\vspace{-1em}
\section{Conclusion}
Predicting ESG ratings from news automatically, without human intervention, could enable various stakeholders, like private investors and government agencies, to monitor the ESG compliance of companies in a time and resource efficient manner. 

Using news articles from over 30,000 different domains and about 200 countries, we have shown that representative ESG ratings can be predicted from information provided in the news. Three steps of input processing are completed to derive the input for the final prediction model. The first and the second step leverage two different BERT models to classify the articles with respect to their ESG relevance and to determine their sentiment. In the third analysis step, the articles are clustered. Comparing different deep learning models for timeseries prediction, the deep convolutional neural network, consisting of three convolutional blocks, provides the best performance. This model outperforms all provided baselines when comparing the average absolute difference to the Asset4 ESG ratings. Further the results imply that regression models fit the task of ESG rating prediction better.
Results show that the derived cluster-based timeseries have a significant impact on the performance of the ESG prediction and thus are an important feature to be considered. We observe also better performance of the models on small cap companies than on larger ones.

In future, it would be interesting to see how results change when a larger company dataset like one that includes 10,000 or more companies is used. 
Further the analysis could be extended with respect to adding other numerical timeseries as input to the prediction. For instance, the amount of green house emission or similar information could be included. In the future, we also plan to incorporate other data sources related to companies such as data stored in Wikipedia or in financial reports (e.g., 10Q reports) which are being increasingly analyzed by automatic approaches nowadays.

 \bibliographystyle{agsm}
 \bibliography{literature}

\addcontentsline{toc}{section}{Appendix}
\section*{Appendix}

\begin{table}[H]
\centering
\caption{Selected Training Parameters for the Classification Model of ESG-related News Articles}
\label{train_para}
\resizebox{.4\textwidth}{!}{%
\begin{tabular}{llr}
\hline
\textbf{Parameter}               %
& \multicolumn{1}{l}{\textbf{Value}} \\ \hline
Learning Rate - Step 1           
& 10-3                               \\
Learning Rate - Step 2           
& 10-4                               \\
Early Stopping Criteria          
& 1                                  \\
Learning Rate Reduction Criteria %
& 5                                  \\ \hline
Maximum Sequence Length          %
& 512                                \\
Batch Size                       %
& 32                                 \\
Loss Funcion                     & %
Sparse categorical cross-entropy loss     \\
Evaluation Metric                & %
Accuracy                              \\ \hline
\end{tabular}%
}
\end{table}

\begin{table}[H]
\centering
\caption{CNN Model Parameters}
\label{cnn_para}
\resizebox{0.4\textwidth}{!}{%
\begin{tabular}{lrrrr}
                                       & \multicolumn{1}{c}{\textbf{Basic CNN}} & \multicolumn{3}{c}{\textbf{Deep CNN}}                                                                              \\ \cline{2-5} 
\multicolumn{1}{c}{\textbf{Parameter}} & \multicolumn{1}{c}{\textbf{Block 1}}   & \multicolumn{1}{c}{\textbf{Block 1}} & \multicolumn{1}{c}{\textbf{Block 2}} & \multicolumn{1}{c}{\textbf{Block 3}} \\ \hline
Convolutional 1D layer                 & \multicolumn{1}{c}{}                   & \multicolumn{1}{c}{}                 & \multicolumn{1}{l}{}                 & \multicolumn{1}{l}{}                 \\
\multicolumn{1}{r}{Filter Size}        & 64                                     & 32                                   & 64                                   & 128                                  \\
\multicolumn{1}{r}{Kernel Size}        & 2                                      & 3                                    & 2                                    & 1                                    \\
\multicolumn{1}{r}{Padding}            & same                                   & same                                 & same                                 & same                                 \\
Dense Layer: Relu Activation           & 265                                    & -                                    & -                                    & -                                    \\ \hline
\end{tabular}%
}
\end{table}

\begin{table}[H]
\centering
\caption{Transformer Model Parameters}
\label{tf_para}
\resizebox{0.4\textwidth}{!}{%
\begin{tabular}{rrrlr}
\multicolumn{1}{l}{}                           & \multicolumn{2}{c}{\textbf{CNN Part}}                                       &  & \multicolumn{1}{c}{\textbf{Transformer Part}} \\ \cline{2-3} \cline{5-5} 
\multicolumn{1}{c}{\textbf{Parameter}}         & \multicolumn{1}{c}{\textbf{Block 1}} & \multicolumn{1}{c}{\textbf{Block 2}} &  & \multicolumn{1}{c}{\textbf{All Blocks}}       \\ \hline
\multicolumn{1}{l}{Convolutional 1D layer}     & \multicolumn{1}{c}{}                 & \multicolumn{1}{c}{}                 &  & \multicolumn{1}{l}{}                          \\
Filter Size                                    & 64                                   & 128                                  &  &                                               \\
Kernel Size                                    & 3                                    & 1                                    &  &                                               \\
Padding                                        & same                                 & same                                 &  &                                               \\ \hline
\multicolumn{1}{l}{Multi-Head Attention Layer} &                                      &                                      &  &                                               \\
Number of Heads                                &                                      &                                      &  & 8                                             \\
Head Size                                      &                                      &                                      &  & 200                                           \\
Dropout                                        &                                      &                                      &  & 0.2                                           \\ \hline
\end{tabular}%
}
\end{table}

\begin{table}[H]
\centering
\caption{General Model Parameters}
\label{para_all}
\resizebox{0.3\textwidth}{!}{%
\begin{tabular}{lc}
\hline
\multicolumn{1}{c}{\textbf{Parameter}}       & \textbf{Value} \\ \hline
Number   of Epochs                           & 25 / 50            \\
Batch   Size                                 & 16             \\
Optimizer                                    & RAdam          \\
Learning   Rate                              & 1e-3           \\ \hline
                                             &                \\ \hline
Learning Rate Reduction Criteria: patience & 5 epochs       \\
Reduction   Factor                           & 0.1            \\
Early Stopping Criteria: patience            & 5 epochs       \\ \hline
\end{tabular}%
}
\end{table}

\begin{table}[H]
\centering
\small
\caption{Overview of assigned Sentiment Labels}
\label{sent_yearly}
\resizebox{0.33\textwidth}{!}{%
\begin{tabular}{ccllll}
\multicolumn{1}{l}{}            & \textbf{year} & \multicolumn{1}{c}{\textbf{positive}} & \multicolumn{1}{c}{\textbf{negative}} & \multicolumn{1}{c}{\textbf{pos/neg}} & \multicolumn{1}{c}{\textbf{pos-neg}} \\ \hline
\textbf{ESG}   & \textbf{2018} & 78.1\%                                & 21.9\%                                & 3.570                                & 0.562                                \\
                                & \textbf{2019} & 78.7\%                                & 21.3\%                                & 3.687                                & 0.573                                \\
                                & \textbf{2020} & 75.3\%                                & 24.7\%                                & 3.057                                & 0.507                                \\ \hline
\textbf{Noise} & \textbf{2018} & 77.9\%                                & 22.1\%                                & 3.523                                & 0.558                                \\
                                & \textbf{2019} & 77.5\%                                & 22.5\%                                & 3.445                                & 0.550                                \\
                                & \textbf{2020} & 74.9\%                                & 25.1\%                                & 2.987                                & 0.498                                \\ \hline
\end{tabular}%
}
\end{table}

\end{document}